\documentclass{article}
\begin{document}
\title{Gravitational hedgehog, stringy hedgehog and stringy sphere}
\author{\"{O}zg\"{u}r Delice\footnote{Department of Physics, Bo\u{g}azi\c{c}i University, 34342  Bebek, Istanbul,
Turkey; e-mail: odelice@boun.edu.tr}} \maketitle
\begin{abstract}
We investigate the solutions of Einstein equations such that a
hedgehog solution is matched to different exterior or interior
solutions via a spherical shell. In the case where both the
exterior and the interior regions are hedgehog solutions or one of
them is flat, the resulting spherical shell becomes a stringy
shell. We also consider more general matchings and see that in
this case the shell deviates from its stringy character.
\end{abstract}
\maketitle
\section{Introduction}
Topological defects may arise during 
 phase transitions in the early Universe \cite{kibble}. Depending
on the type of the spontaneously broken symmetry, monopoles,
strings, domain walls and textures may be produced
\cite{vilenkin}.
Among these defects, cosmic strings are  studied extensively. They
have wider applications in galaxy formation and gravitational
lensing theories which might have observational implications
\cite{vilenkin}.

There are also some solutions which are constructed from ensembles
of cosmic strings.
 The energy momentum tensor of an infinitely long straight cosmic
string along $z$ axis is given by
\begin{equation}\label{EMTofCS}
T_{zz}=-T_{00}=-\mu\, \delta(x)\delta(y),
\end{equation}
with other components vanishing \cite{VilenkinCS}. The spacetime
outside the straight cosmic string is flat but conical since the
well known exterior metric of an infinitely long cosmic string
contains an angular deficit. Namely in usual cylindrical
coordinates:
\begin{equation}\label{CSmetric}
ds^2=-dt^2+dr^2+dz^2+(1-4 G \mu)^2r^2d\phi^2,
\end{equation}
where the angular coordinate $\phi$ has the range $0\le \phi \le
2\pi$ and $\mu$ is the mass per unit length of the string.

Since we will present a solution with an ensemble of cosmic
strings, here we  discuss some 
solutions of the Einstein equations giving rise to cosmic stringy
structures. By cosmic stringy structures we mean a $d$ dimensional
hypersurface in 4 dimensional space-time ($d\leq 3$) such that on
this
hypersurface the principal stresses 
of the
Energy-Momentum tensor satisfy
\begin{equation}\label{condition}
\sum_{i=1}^d p_i=-\rho,
\end{equation}
where $\rho$ is the energy density and $p_i$ are the principal
stresses. The famous solution of Vilenkin \cite{VilenkinCS}
describing an infinitely long straight cosmic string is our first
example of these structures where the string is taken as a
$\delta$-function source ($d=1$). On the string condition
(\ref{condition}) is satisfied with $p_z=-\rho$.

One example of the two dimensional stringy structures satisfying
 condition (\ref{condition}) is the stringy hollow cylinder
solution where an infinitely long cylinder is constructed from
straight strings \cite{Tsoubelis}. In this solution, the interior
region is flat and the exterior region is  given by a locally flat
cosmic string metric (\ref{CSmetric}). This solution can be
extended to the rotating hollow stringy cylinder case
\cite{Clement}. The solution we will present in section
(\ref{Stringy sphere}) is another example of cosmic stringy
surfaces where  strings lie on the surface of a sphere. Unlike the
hollow cylinder case, either both or the interior or the exterior
regions are not vacuum but contain radial strings.

The solutions with multiple parallel or nonparallel straight
cosmic strings moving with different velocities are given in
\cite{Letelier}. Thus using this solution, one can orient
strings even non parallel. Using  a different method,
 Arik et. al. presented a solution \cite{Arik} where a
straight cosmic string decay into conical surfaces at its end
points and ending up as a stringy sheet with radial strings. This
solution can be extended to the other cases \cite{delice}
including a string changing into a cylinder, a cylinder changing
into a cylinder with different radius and a cylinder decaying into
conical surfaces at its endpoints and ending up as a sheet with
radial strings.

The Gott-Hischok-Linet interior solutions \cite{Gott}-\cite{Linet}
are three dimensional examples of the cosmic stringy structures
where a solid cylinder is filled with long string
fibers with $p_z=-\rho$. In the first two solutions 
 the energy density is constant in the cylinder whereas in the third one
it is  variable. Both solutions match to the string exterior
smoothly. These are the first examples of stringy volumes.

One of the earlier solutions of the stringy volumes is given by
Letelier \cite{Letelier1}. The solution has a cloud of strings as
a source lying between the Schwarzschild exterior and a fluid
interior. The line element given in this solution also describes a
global monopole (hedgehog) which we will review in the next
section.

 The coasting universe \cite{Kolb} is a cosmological solution where
 the universe is dominated by cosmic strings \cite{VilenkinK}. The equation of
state of the fluid which fills up this universe is $p=-1/3\,\rho$.
This can be thought to be made of randomly oriented straight
strings.

Recently, another example of the stringy volumes and surfaces has
been given in \cite{Kirsch}. This is a solution with an interior
with a string-like equation of state, a surface layer with again a
string-like equation of state and a locally AdS exterior in
toroidal coordinates.

 Unlike straight strings, curved strings have nonzero
gravitational potential \cite{vilenkin}. Hence, they cannot be
static. It is known that a loop of string may oscillate or, due to
its large tension, either emits radiation and vanishes completely,
or collapses and emits some portion of its energy as radiation and
forms a black hole \cite{HawkingCs}. To have a loop stable one
needs to support it again collapse. A circular string solution is
given in \cite{Frolov} as an initial value problem. The solution
we present in section (\ref{Stringy sphere}) contains strings
lying on a surface of a sphere. In our solution these strings are
supported by radial strings.

Actually, there could be more solutions that we are not aware of.
However, the solutions we have mentioned reflect the richness of
the situation of constructing more complex structures such as
stringy surfaces or volumes using cosmic strings.

In this paper we will mainly use solutions of Einstein equations
with two different interpretations. 
The results we will find might have some applications to the
theories ( String theory, GUT, Superstrings ...) which can produce
defects. In the next section, we will first review the
gravitational field of a gravitational hedgehog solution
\cite{BariolaVilenkin} arising from  triplet of scalar fields.
Then following \cite{Guendelman}, we see that the same
gravitational field can also be constructed using an ensemble of
straight radial cosmic strings intersecting at a central point.
This solution is called the "string hedgehog" solution. To better
understand such spacetimes, in section (\ref{Stringy sphere}) and
(\ref{Generalization}) we study possible matchings of this
space-time to different interior or exterior space-times. Our main
observation will be that when we match two different hedgehog
solutions via a spherical shell, a stringy-sphere arises.
 We will also discuss under which conditions the shell, the exterior and interior
regions satisfy some energy conditions. In the last section we
perform more general matchings and realize that the shell will deviate from its stringy character.

\section{Gravitational hedgehog and string hedgehog}
\subsection{Gravitational field of a global monopole (hedgehog)}
The simplest model  that gives rise to the hedgehog
\cite{BariolaVilenkin} is described by the Langrangian
\begin{equation}
L=\frac{1}{2}\partial_\mu\phi^a\partial^\mu\phi^a-\frac{1}{4}\lambda
\left( \phi^a \phi^a-\eta^2\right)^2,
\end{equation}
where $\phi^a$ is a triplet of scalar fields, $a=1,2,3$ . The
model has a global O(3) symmetry, which is spontaneously broken to
U(1). The field configuration describing a monopole is

\begin{equation}\label{scalarfields}
\phi^a=\eta f(r) x^a/r
\end{equation}
where $x^a x^a=r^2$. In flat space the hedgehog core has size
$\delta\sim \lambda^{-1/2}\eta^{-1}$ and mass $M_{core}\sim
\lambda^{-1/2}\eta .$ Folloving \cite{BariolaVilenkin} we can take
$f(r)=1$ outside the core and the energy momentum tensor becomes
\begin{equation}\label{EMT}
T_{rr}\approx-T_{00}\approx-\eta^2/r^2, \quad
T_{\theta\theta}\approx T_{\phi\phi}\approx 0.
\end{equation}

Let us consider the general metric
\begin{equation}\label{generalmetric}
ds^2=-A^2(r)dt^2+B^2(r)dr^2+C^2(r)d\Omega_2^2.
\end{equation}
Choosing the orthonormal basis one forms as
\begin{equation}
e^0=A(r)dt, e^1=B(r)dr, e^2=C(r)d\theta, e^3=C(r)\sin\theta d\phi,
\end{equation}
 for the nonzero components
of the Einstein tensor for this metric  one finds:
\begin{eqnarray}
G_{rr}&=&\frac{2A_rC_r}{AB^2C}-\frac{1}{C^2}\left(1-\frac{C_r^2}{B^2}
\right),\\
G_{\theta\theta}&=&\frac{A_{rr}}{AB^2}+\frac{C_{rr}}{B^2C}-\frac{A_rB_r}{AB^3}+\frac{A_rC_r}{AB^2C}-\frac{B_rC_r}{B^3C} \\
&=&G_{\phi\phi},\\
G_{00}&=&-\frac{2C_{rr}}{B^2C}+\frac{2B_rC_r}{B^3C}+\frac{1}{C^2}\left(1-\frac{C_r^2}{B^2}\right).
\end{eqnarray}
An immediate solution of (\ref{EMT}) is found by choosing
$A(r)=B(r)=const.$, $C(r)=\alpha r$. Then the metric becomes

\begin{equation}\label{metric1}
ds^2=-dt^2+dr^2+\alpha^2r^2d\Omega_2^2, \quad
d\Omega_2=d\theta^2+r^2\theta d\phi^2,
\end{equation}
with the only nonzero components of the Einstein tensor satisfying

\[
G_{00}=-G_{rr}=\frac{1-\alpha^2}{\alpha^2r^2}.
\]
This solution describes  the asymptotic behavior of the exterior
field of a global monopole outside the core and is first presented
in \cite{Starobinsky} when studying spacetimes with angular
deficit. Also, unlike straight cosmic strings, the exterior field
of a gravitational monopole is not flat and it contains a solid
deficit angle.

Notice that the metric (\ref{metric1}) can be put into isotropic
form
\[
ds^2=-dt^2+(x^2+y^2+z^2)^{1-\beta}(dx^2+dy^2+dz^2)
\]
with
\[
\rho=\alpha r, \quad \beta=\alpha^{-1}, \quad
 (x,y,z)=\rho^{\beta}(\sin\theta \sin\phi,\sin\theta \cos\phi,
cos\theta).
\]

 The general solution of the metric (\ref{generalmetric}) satisfying (\ref{EMT}) with
 $C(r)=r$ is given by \cite{Letelier}
\begin{equation}\label{metric2}
A(r)=B^{-1}(r)=\left(1-8\pi G \eta^2- \frac{2Gm}{r^2}  \right).
\end{equation}
Here $m$ is a constant of integration and $m\sim M_{core}$
\cite{BariolaVilenkin}. This solution is not asymptotically flat,
see \cite{Ulises1} for discussion of its ADM mass. It is shown in
\cite{Harari} and \cite{shi} that $m $ may be negative and is
responsible for tiny repulsive force of the core of the hedgehog.
The dynamics and the stability of its core is discussed in
\cite{Cho1,Cho2}. Notice that for large and positive $m$, the
solution corresponds to a hedgehog swallowed by a black hole
\cite{BariolaVilenkin, Ulises2}. Actually, we can neglect the mass
term $m$ for reasonable values of $\eta$ and $\lambda$ on the
astrophysical scale \cite{BariolaVilenkin}. When $m=0$, the
general solution (\ref{metric2})  reduces to (\ref{metric1}) after
rescaling $r$ and $t$ coordinates with $\alpha^2=(1-8\pi G
\eta^2).$

Thus so far we have a triplet of scalar fields $\phi^a$ of the
form (\ref{scalarfields}) generating gravitational field around
the hedgehog (monopole). As shown in \cite{Guendelman}, the
gravitational field around the hedgehog can also be constructed
using an ensemble of radial cosmic strings.

\subsection{Gravitational field of a string hedgehog}

Let us suppose that we have an ensemble of radially oriented
cosmic strings, whose gravitational field is given in
(\ref{EMTofCS}), all of them intersecting at a central point. If
we take the strength of each string to be very small but the
number of strings very big, then the resulting configuration will
approximately have spherical symmetry. In the continuum limit,
this symmetry will be exact. This is called "string hedgehog"
\cite{Guendelman}. The nonzero components of the energy-momentum
tensor of this configuration will satisfy
\begin{equation}\label{EMTofSH}
T_{rr}=-T_{00}.
\end{equation}

In this and the next section we will take $m=0$. Thus, we will use
the metric (\ref{metric1}). Note that the energy contained inside
a sphere of radius $r_0$ is given by
\begin{equation}
E=\int_0^{r_0}T_{00}\ e^1\wedge e^2 \wedge e^3=4 \pi
(1-\alpha^2)r_0,
\end{equation}
where$ \ e^1=dr,\ e^2=\alpha r d\theta,\ e^3= \alpha r \sin \theta d
\phi$, and is
 linear in the proper radius of
this sphere.

It is shown  that we can have solutions with radial strings
intersecting from a common point with \cite{Aryal, Ivanov} or
without \cite{Dowker, FrolovFursaev} a black hole at the center.
It is also mentioned in those solutions that if mass per unit
length of the strings (which is proportional to their angular
deficit) is very small then we can go to the continuum limit. This
actually corresponds to the situation we consider in this section.

\section{Stringy hedgeball\label{Stringy sphere}}

In this section we match this string hedgehog solution to another
hedgehog solution with the parameter $\beta$. When $\alpha=1$ or
$\beta=1$, the interior or the exterior regions have flat
Minkowski metric.
Taking the  exterior region  Minkowskian may sound unphysical,
but, if we only consider the limits where $m=0$ and $\alpha^2
\approx 1 \quad (\eta^2 <<1)$  than for large $r$ the exterior
region becomes almost Minkowskian. Thus, we consider the solution
with Minkowski exterior  valid in these limits.

  As we have emphasized before, for the string hedgehog solution,
the energy density and the radial pressure $\rho_r$ is vanishing
at $\infty$. In this section we will try to find out the answer of
this question: Is it possible to match this spacetime to
another string hedgehog solution with different parameter such
that both the interior and the exterior regions and also the shell
satisfy certain energy conditions?
Since for exterior and interior regions we will have different
radial pressures, from continuity, we cannot smoothly match these
two regions. Thus, we need a surface layer (infinitely thin shell)
at the boundary of these two regions.

 We take the interior metric as (\ref{metric1}) and the exterior
metric as

\[
ds^2_+=-dt^2+d\rho^2+\beta^2 \rho^2d\Omega^2_2.
\]
There are several methods to calculate the energy momentum tensor
of the shell \cite{israel2shell}-\cite{Mansouri}. Here we use an
alternative method
\cite{Mansouri} which is equivalent to the
standard method presented by Israel\cite{israel2shell}.
We also follow Lichnerowitz boundary
conditions for surface layers \cite{Lichnerowicz}.
 The metrics interior and exterior to the hypersurface seperating
 these two regions must be continuous everywhere, but their derivatives
with respect to the radial coordinate may contain discontinuities
which give rise to the energy momentum tensor of the shell.
 This is the
condition to have a surface layer at $r=r_0$. To do this we choose
$\rho=\rho(r)$ and using the boundary conditions given below we
can make the metric continuous at $r=r_0$.  From the continuity
conditions of the metric, at $r=r_0$ we have:
\[
\rho(r_0)=\frac{\alpha}{\beta}\, r_0; \quad \rho_{,r}(r_0)=1.
\]
Choosing $\rho(r)=ar+b$ and using boundary conditions one gets

\[
\rho(r)=r+(\alpha/\beta -1)r_0.
\]
Thus the exterior and interior metrics become
\begin{eqnarray}\label{stringMinkowskimetr}
ds^2_+&=&-dt^2+dr^2+\beta^2\left(r+(\alpha/\beta-1)r_0 \right)^2d\Omega_2^2\\
ds^2_-&=&-dt^2+dr^2+\alpha^2r^2d\Omega_2^2.
\end{eqnarray}
With the help of the Heaviside step function we can combine both
the interior and the exterior metrics in the form
\[ds^2=\theta(r-r_0)ds^2_++\theta(r_0-r)ds^2_-.\]
After calculating the Einstein tensor for this metric, the terms
proportional to Dirac delta function will give the energy momentum
tensor of the shell and the terms proportional to the step
functions give the interior and the exterior solutions. Thus, the
energy-momentum tensor for the whole spacetime can be expressed as
\begin{equation}
T_{\mu\nu}= T_{\mu\nu}^{(-)}\, \theta(r_0-r)+T_{\mu\nu}^{(+)}\,
\theta(r-r_0)+ {T}_{\mu\nu}^{(0)}\, \delta(r-r_0),
\end{equation}
where
\begin{equation}\label{Diagonalemt}
T^{(k)}_{\mu\nu}=diag\left(\rho^{(k)},p_i^{(k)}\right),\quad
k=\pm,0; \quad i=r,\theta, \phi.
\end{equation}
Calculating the Einstein tensor, one gets for the nonzero
components
\begin{eqnarray}\label{EmtofExtMinIntMon}
\rho^{(-)}&=&-p_r^{\phantom{A}(-)}=\frac{1-\alpha^2}{\alpha^2r^2},
\\
\rho^{(+)}&=&-p_r^{(+)}=\frac{1-\beta^2}{\beta^2r^2}, \\
\rho^{(0)}&=&\frac{2(\alpha-\beta)}{\alpha r}, \quad
p_\theta^{(0)}=p_\phi^{(0)}=\frac{(\beta-\alpha)}{\alpha r}.
\end{eqnarray}

 So, we have matched a string hedgehog space-time to another string
 hedgehog solution via a surface layer at the boundary.
  Since we do not want to discuss the solutions where
  $\alpha^2=1-8\pi G  \eta^2$ is negative, we limit the range of
   the parameters $\alpha$ and $\beta$ as $-1\leq\{\alpha,\beta\}\leq1$ since we
 have
 $0<\{\alpha^2, \beta^2\}\leq1$.

For an energy momentum tensor of the form (\ref{Diagonalemt}) we
have the weak energy condition ($\rho\geq0, p_i\geq0$), the
dominant energy condition ($\rho\geq|p_i|$) and strong energy
condition ($\rho+\sum_i p_i\geq0$) \cite{HawkingEllis}. Thus, the
weak energy condition is not satisfied since $p_i$ is negative.
When $\alpha$ and $\beta$ have the same signs,  to have a positive
energy density, we need $|\alpha|>|\beta|$. When they have
different signs the energy density is always positive. The shell
has the equation
 of state:
 \begin{equation}\label{EOstringyshell}
\rho^{(0)}=-\left(p_\theta^{(0)}+p_\phi^{(0)}
 \right).
 \end{equation}
Then, when the shell has positive energy density, it satisfies
dominant and strong energy conditions. Note that both the interior
and the exterior regions also satisfy these energy conditions for
these ranges of the parameters. The total energy of the shell is
$E=8\pi \alpha(\alpha-\beta)r_0$.

  We can interpret this shell as a shell composed of uniformly distributed
  cosmic strings lying on the surface of a sphere since it satisfies (\ref{condition}).

 Thus when we want to embed a stringy hedgehog spacetime to another hedgehog solution,
 a stringy spherical shell arises. For a certain range of the parameters, both the shell
 and the  string hedgehog spacetime satisfies energy conditions.

If we take $\beta=1$ the exterior region becomes flat Minkowski
spacetime. In this case to have a shell satisfying energy
conditions, we need $-1<\alpha<0$. We can call this solution with
flat exterior a "stringy hedgeball"  solution. Or, we can discuss
the opposite situation. If we take $\alpha=1$, then the interior
region of the spherical shell becomes flat. For this case, to
satisfy energy conditions, one needs $0<\beta<1$.

\section{Generalizations and discussions\label{Generalization}}

In this section we consider the interior and the exterior metrics
$ds^2_{\pm}$ as hedgehog-Schwarzschild-de Sitter solutions since
we can superpose the Schwarzschild- de Sitter solution with the
hedgehog solution \cite{Guendelman}. For the interior and the
exterior regions we label the coordinates as $(t,r,\theta,\phi)$
and $(\tau, R, \theta, \phi)$. The metrics are of the form
(\ref{generalmetric}) where the metric functions are given as
\begin{eqnarray}
A_-(r)&=&B_-^{-1}(r)=\left(\alpha^2-\frac{2m}{r}-\frac{\Lambda_-}{3}r^2\right)^{1/2},
 \\
A_+(R)&=&B_+^{-1}(R)=\left(\beta^2-\frac{2M}{R}-\frac{\Lambda_+}{3}\,
R^2 \right)^{1/2}, \\
  C_-(r)&=&r,\  C_+(R)=R .
\end{eqnarray}
We require them to be continuous at $r=r_0$. We again keep the
interior metric as it is and for the exterior metric  we choose
$R=R(r)=ar+b$ and $\tau=Et$. Then we have the following boundary
conditions:
\begin{equation}
R(r_0)=r_0, \quad R_{,r}(r_0)=\frac{A_+(r_0)}{A_-(r_0)}, \quad
E=\frac{A_-(r_0)}{A_+(r_0)}.
\end{equation}
Our exterior metric becomes
\begin{equation}
ds^2_+=-A_+^2(r)dt^2+A_+^{-2}(r)dt^2+R^2(r)d\Omega_2^2.
\end{equation}
with
\begin{eqnarray}
A_+(r)=\sqrt{\beta^2-\frac{2M}{R(r)}-\frac{\Lambda_+R^2(r)}{3}
}\frac{A_-(r_0)}{A_+(r_0)}, \\ R(r)=\frac{A_+(r_0)}{A_-(r_0)}\, r
+\left(1-\frac{A_+(r_0)}{A_-(r_0)}\right)r_0.
\end{eqnarray}
The energy momentum tensor of the shell has the following
components
\begin{equation}
\rho^{(0)}=2K(r_0), \quad
p_\theta^{(0)}=p_\phi^{(0)}=-K(r_0)+L(r_0),
\end{equation}
where
\begin{eqnarray}
K(r_0)&=&\frac{A_-(r_0)}{r_0}\big(A_-(r_0)-A_+(r_0) \big), \phantom{AAAAAAAAAAAA}\\
 L(r_0)&=&-\frac{A_-(r_0)}{A_+(r_0)}\left(\
\frac{\Lambda_+r_0}{3}-\frac{M}{r_0^2}\right)+\left(\frac{\Lambda_-r_0}{3}-\frac{m}{r_0^2}
\right).\phantom{AAAA}
\end{eqnarray}

This solution is quite general since choosing some parameters zero
we can recover the more simple solutions such as when
$m,M,\Lambda_\pm$ vanishing, the solution reduces to the stringy
shell solution we have presented in the previous section. If
$\Lambda_\pm,=0, \alpha=\beta=1 $ we have a shell around a black
hole\cite{Frauendiener}.

Note that this shell no longer satisfies (\ref{EOstringyshell}).
However, whenever the parameters other than $\alpha, \beta$ go to
zero, $L(r_0)$ also goes to zero and we recover
(\ref{EOstringyshell}). The presence and the difference of these
parameters deviates the shell from its stringy character.  For
this general case we can specify an equation of state for the
shell $p_i=p_i(\rho)$ and solve for the parameters since for this
case we have more parameters than the equations. We can even have
a domain wall satisfying $p^{(0)}_\theta=p^{(0)}_\phi=-\rho^{(0)}$
if the parameters satisfy $K(r_0)+L(r_0)=0$ or a shell of photons
satisfying $T^{\mu(0)}_{\phantom{\mu}\mu}=0$ if the relation
$4K(r_0)-L(r_0)=0$ holds.

In this paper using the thin shell formalism, we have glued some
interior solutions in spherical coordinates to an exterior
solution via a spherical shell where the interior and the exterior
solutions have the same character with different parameters.

 After reviewing the global hedgehog (monopole) and the string hedgehog
solutions, we have first chosen the interior and the exterior
regions as the string hedgehog solution which can be made up with
an ensemble of  radial cosmic strings and we have calculated the
energy momentum tensor of the shell seperating these two regions.
We see that this infinitely thin shell is composed of cosmic
strings lying on the surface of the sphere. For certain ranges of
the parameters, the interior, the exterior and  the shell satisfy
the dominant energy condition.

 Then we have chosen the interior and the exterior regions
as Hedgehog-Schwarzschild-de Sitter solutions  with different
parameters, and we have seen that for this case the shell is no
longer stringy.

\section*{Acknowlegment}
 I would like to thank M. Ar\i k and T. Turgut
for reading the manuscript and useful discussions.

\end{document}